\documentclass{iaus}
\usepackage{graphicx}

\title[Stellar Populations] 
{Stellar Populations in Barred Galaxies}

\author[Robert, Cantin, Moll\'a, Pellerin, \& Bri\`ere]   
{C. Robert$^1$, S. Cantin$^1$, M. Moll\'a$^2$, A. Pellerin$^3$,
 \and \'E. Bri\`ere$^1$}

\affiliation{$^1$Universit\'e Laval \& CRAQ, Qu\'ebec, QC G1V 0A6, Canada, {\tt carobert@phy.ulaval.ca} 
\\[\affilskip]
$^2$Departamento de Investigaci\'on B\'asica, CIEMAT Avda. Complutense 22, 28040 Madrid, Spain
\\[\affilskip]
$^3$Texas A\&M University, Dept of Physics \& Astronomy, College Station, TX 77843, USA
}

\pubyear{2008}
\volume{262}  
\pagerange{119--126}
\jname{Stellar Populations -- Planning for the Next Decade}
\editors{G. R. Bruzual \& S. Charlot, eds.}
\begin{document}

\maketitle

\begin{abstract}
We developed an iterative technique to better characterize stellar populations and the central activity 
of barred  galaxies using evolutionary synthesis codes and OASIS data. 
The case of NGC\,5430 is presented here. Our results are reinforcing the role played by the bar 
and nuclear structures for the evolution of galaxies.

\end{abstract}

\firstsection 
\firstsection
\section{Introduction}

The assemblage of the components of a galaxy is a puzzling but important 
subject.  Among the key elements, 
a large scale bar is believed to be  an efficient way to drive gas toward the central kiloparsecs 
of the galaxy (\cite{norman}). This may trigger star formation or turn on the central 
engine (\cite{haan}). Interactions, mergers, or cosmic filaments are considered to play a role in the accretion 
of gas and its flow process as well (\cite{combes}). One way to deepen our understanding 
of these phenomena is to study in details the content of galaxies.

We used the spectro-imager OASIS at the CFHT and WHT to describe the stellar populations 
in the central region of 8 galaxies. We developed an iterative technique 
to separate the flux from two stellar populations, a young population ($<$15\,Myr) and an old 
underlying one (see Cantin et al. 2009 for more details).  In summary, 
the young population is characterized  by comparing 
the equivalent width of H$\alpha$ and H$\beta$ with starburst models from LavalSB (\cite{dionne}). 
The old population is studied using absorption lines (e.g. Mg$_2$, Fe{\sc{i}}) and models of \cite{gonz}.
The iterative process takes into account the flux and line indicators from one population 
while studying the other one. It allows us, for example, to isolate the absorption 
component in H$\beta$  and therefore get more accurate  age and mass estimates. 
Reliable uncertainties,  within the models considered, are obtained using Bayesian statistics.

\firstsection
\section{NGC\,5430}

\begin{figure}[t]
\begin{center}
 \includegraphics[width=5.3in]{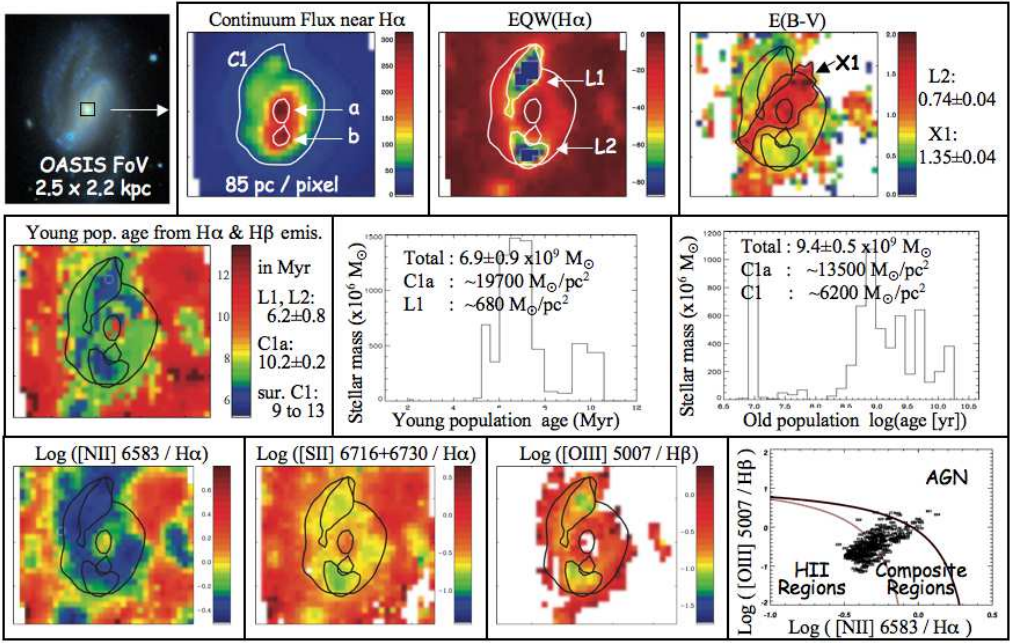} 
\caption{OASIS data for the central region of NGC\,5430.
Contours C1, a, and b refer to the continuum flux, 
contours L1 and L2 to the emission lines, and X1 to the extinction.}
\label{fig1}
\end{center}
\end{figure}

NGC\,5430 is a nearby (42\,Mpc) SB(s)b starburst galaxy. While continuum
flux images show a double peak structure, a nuclear spiral/ring is clearly seen from  the emission 
lines (Fig.\,\ref{fig1}). The extinction map indicates a dust lane located  between the emission knots, 
and oriented along the galaxy large scale bar. With the iteration technique we find a gas metallicity 
close to 3\,Z$_\odot$. The emission regions L1 and L2 are only 6.2$\pm$0.8\,Myr.
The old population distribution is more uniform. 
It was  formed over a long period of time, between 700\,Myr and 4\,Gyr. It has a much lower metallicity 
(0.4$\pm$0.2\,Z$_\odot$ on average) than the gas and is barely dominating the stellar mass. 
The regions L1 and L2 are typical H{\sc{ii}} regions based on the emission line ratios. 
A weak signature from a Seyfert or LINER  is possibly present in the galaxy nucleus. 
Emission typical of composite regions (\cite{kewley}) is also found far from the nucleus. 
New data obtained with SpIOMM (a Fourier transform spectro-imager; \cite{drissen}) 
is confirming a high ratio  [N{\sc{ii}}]/H$\alpha$  in the galaxy disk 
and is showing many H{\sc{ii}} regions along the galaxy bar with an age near 7\,Myr.

\firstsection
\section{Conclusions}

For most galaxies in our sample, we find:  1) star forming knots of age 5-10 Myr distributed in a nuclear bar, 
spiral, or ring; 2) a peculiar dust distribution; 3) an underlying old (0.1-5 Gyr) generation of stars
with a metallicity lower than the gas; and 4) in most cases, a non thermal activity, 
either from the composite or the LINER type and not necessarily restricted to the nucleus.
We  see relations between the star forming knots morphology, the dust distribution, the variation in the 
gas and stellar populations abundance, along with the galaxy large scale bar orientation. These relations 
confirm the importance of the bar for the galaxy evolution. Furthermore in late type spirals, secular evolution, 
as defined by \cite{kk},  has been proposed to rearrange the material from the disk and build 
up a pseudobulge. It has been described as a slow process, either internal  or environmental.
This could be a simple scenario, involving here the galaxy bar, 
to explain the different stellar populations seen in our study.

\firstsection

\end{document}